\documentclass[12pt]{iopart}

\usepackage{graphicx}
\usepackage{textcomp}
\usepackage{siunitx}
\usepackage{xcolor}
\begin{document}

\title[Stable Branched Electron Flow]{Stable Branched Electron Flow}

\author{B.~A.~Braem, C.~Gold, S.~Hennel, M.~R\"o\"osli, M.~Berl, W.~Dietsche, W.~Wegscheider, K.~Ensslin, T.~Ihn}

\address{ETH Z\"urich, Solid State Physics Laboratory, Otto-Stern-Weg 1, 8093 Z\"urich, Switzerland}
\ead{bbraem@phys.ethz.ch}
\vspace{10pt}
\begin{indented}
\item[]\today
\end{indented}

\begin{abstract}
The pattern of branched electron flow revealed by scanning gate microscopy shows the distribution of ballistic electron trajectories. The details of the pattern are determined by the correlated potential of remote dopants with an amplitude far below the Fermi energy. We find that the pattern persists even if the electron density is significantly reduced such that the change in Fermi energy exceeds the background potential amplitude. The branch pattern is robust against changes in charge carrier density, but not against changes in the background potential caused by additional illumination of the sample.
\end{abstract}

%
%

%
%

\section{Introduction to branched electron flow and scanning gate microscopy}
Semiconductor heterostructures of high purity allow electrons to move ballistically in a smooth potential landscape. This is crucial to achieve high charge carrier mobilities. At the same time investigating charge transport in the buried electron gas on a microscopic level is cumbersome. Therefore little is known about the electrons microscopic behavior and there are surprises like the strongly viscous behavior of charge carriers in high-mobility electron systems \cite{bandurin_negative_2016, moll_evidence_2016, crossno_observation_2016, levinson_viscous_2018}. 

Non-invasive local measurement techniques such as scanning SQUID microscopy \cite{koshnick_terraced_2008,vasyukov_scanning_2013} are limited in their spatial resolution by the separation between the SQUID and the two-dimensional electron gas (2DEG) of the order of 100~nm. Higher resolution can be achieved by the technique of scanning gate microscopy (SGM) \cite{eriksson_cryogenic_1996}. This invasive method creates a movable local potential hill in the plane of the 2DEG. High resolution is achieved when electrons emanating from a quantum point contact (QPC) scatter from this barrier back through the constriction. Topinka \textit{et al.} used this technique to image branched electron flow in a AlGaAs heterostructure \cite{topinka_imaging_2000,topinka_coherent_2001}. These measurements are interpreted to reflect the spatial distribution of electron flow in the unperturbed case \cite{topinka_imaging_2000,heller_branching_2003}.

The anisotropic pattern is caused by the background potential generated by remote ionized donor atoms. Therefore one expects the branch pattern to change if the kinetic energy of the charge carriers changes by the average amplitude of the background potential. In contrast to this expectation our SGM experiments at different electron densities show that the pattern of branched flow is robust. Changes of the Fermi energy up to a factor of two change the branches visibility, but not their position.

\section{Experimental realization}
\begin{figure}
\begin{center}
 \includegraphics[width=\linewidth]{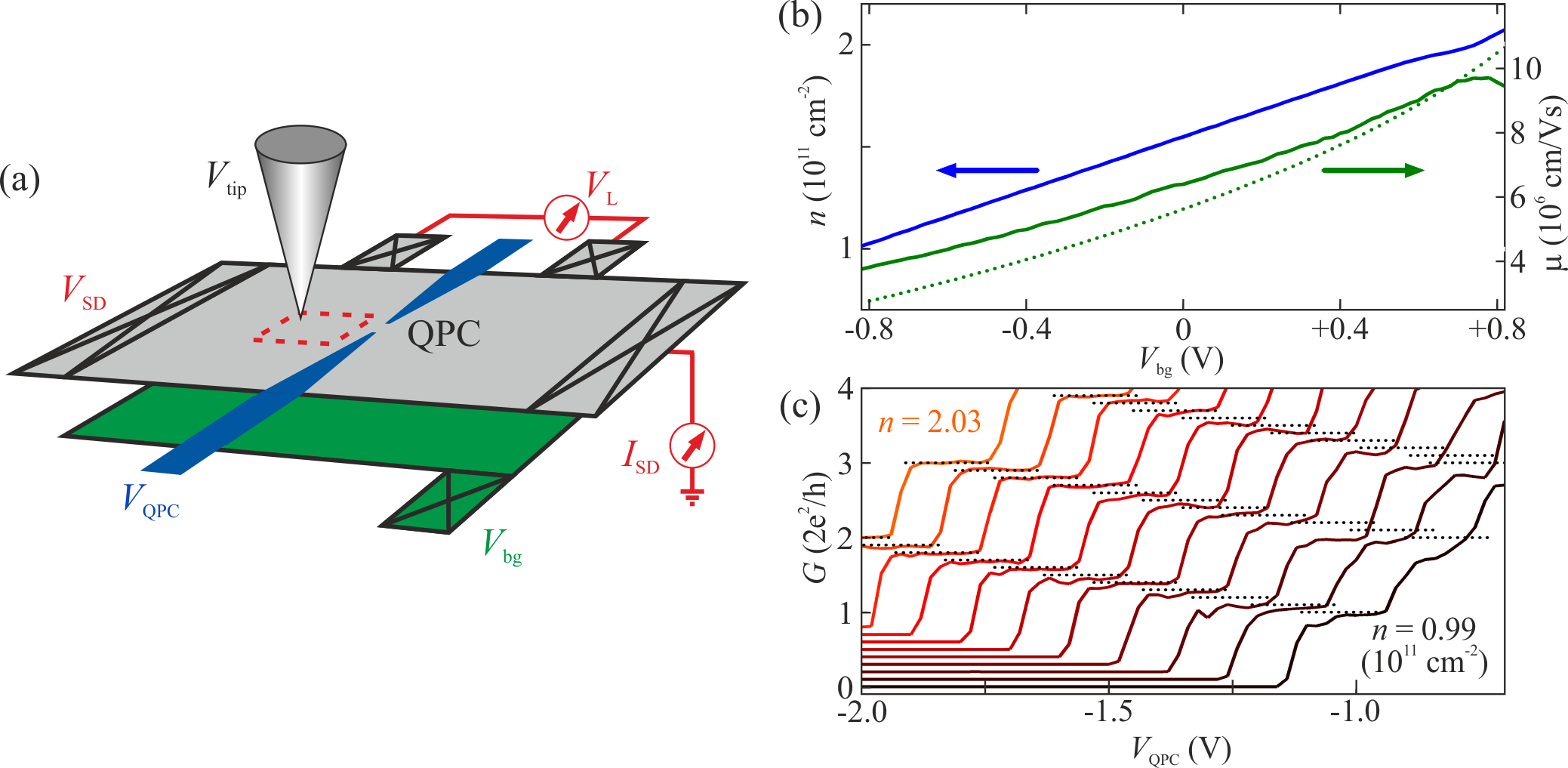}
 \caption{(a) Schematic of the Hall bar shaped sample (grey) with the back-gate (green). The voltage biased AFM tip is scanned 30~nm above the GaAs surface in the scan frame (red dashed) close to the split-gate defined QPC (blue). Longitudinal voltage $V_{\mathrm{L}}$ and source-drain current $I_{\mathrm{SD}}$ are measured as function of tip position. (b) Hall density (blue solid line) and mobility (green) of the illuminated 2DEG as function of back-gate voltage measured in our SGM setup at an electronic temperature $<30$~mK. The green dotted line shows the electron mobility of our model (see section~3). (c) QPC conductance plateaus as a function of split-gate voltage for equidistant electron densities. The voltage biased SGM tip is placed in the center of the scan frame 4~\si{\micro\meter} from the QPC gap. Curves are vertically offset by 0.1$\times2e^2/h$ for clarity, the expected conductance values are indicated by the dotted lines.}
 \label{fig1label}
 \end{center}
\end{figure}

As shown in Fig.~\ref{fig1label}(a), the 2DEG of our GaAs/AlGaAs sample is etched into a Hall bar shape that allows for measuring the longitudinal voltage $V_{\mathrm{L}}$ and the source-drain current $I_{\mathrm{SD}}$ in a four-terminal configuration. The 2DEG is buried 130~nm under the sample surface and separated by $1.13$~\si{\micro\meter} from the back-gate \cite{berl_structured_2016}. We illuminated the sample with red light to increase the charge carrier mobility by ionizing additional donor atoms, which changes the random background potential. This so-called persistent photo conductivity in AlGaAs heterostructures is a well established effect and has been used to tune density and mobility of 2DEGs. After illumination, we can change the electron density $n$ by applying a back-gate voltage $V_{\mathrm{bg}}$ as shown by the blue curve in Fig.~\ref{fig1label}(b). In the presented measurements, we use the back-gate to tune $n$ in the range $1.0-2.0\times 10^{11} \, \mathrm{cm}^{-2}$ where the electron mobility $\mu$ changes in the range $3-8\times 10^{6} \, \mathrm{cm}^2/\mathrm{Vs}$.

The QPC is formed by applying a gate voltage $V_{\mathrm{QPC}}$ to split gates with a lithographic gap of 400~nm. We observe conductance quantization in the entire range of charge carrier density as presented in Fig.~\ref{fig1label}(c). The dips in the conductance plateaus are caused by the voltage biased SGM tip that is placed \SI{4}{\micro\meter} from the QPC. At low charge carrier densities the second and third plateau are tilted and below the expected conductance values indicated by black dotted lines. To remain on the second plateau for different charge carrier densities, we apply a $V_{\mathrm{bg}}$-dependent split gate voltage $V_{\mathrm{QPC}}=V_\mathrm{bg}^2/\SI{24.5}{V} - 0.62\times V_{\mathrm{bg}} - \SI{1.33}{V}$ in the following measurements. 

We measure in a home-built atomic force microscope (AFM) in a dilution refrigerator with a base temperature of 25~mK and an electronic temperature of the sample below 30~mK. To create the movable potential perturbation in the sample we apply a voltage $V_{\mathrm{tip}}$ to the metallic tip.

\begin{figure}[h]
\begin{center}
 \includegraphics[width=0.85\linewidth]{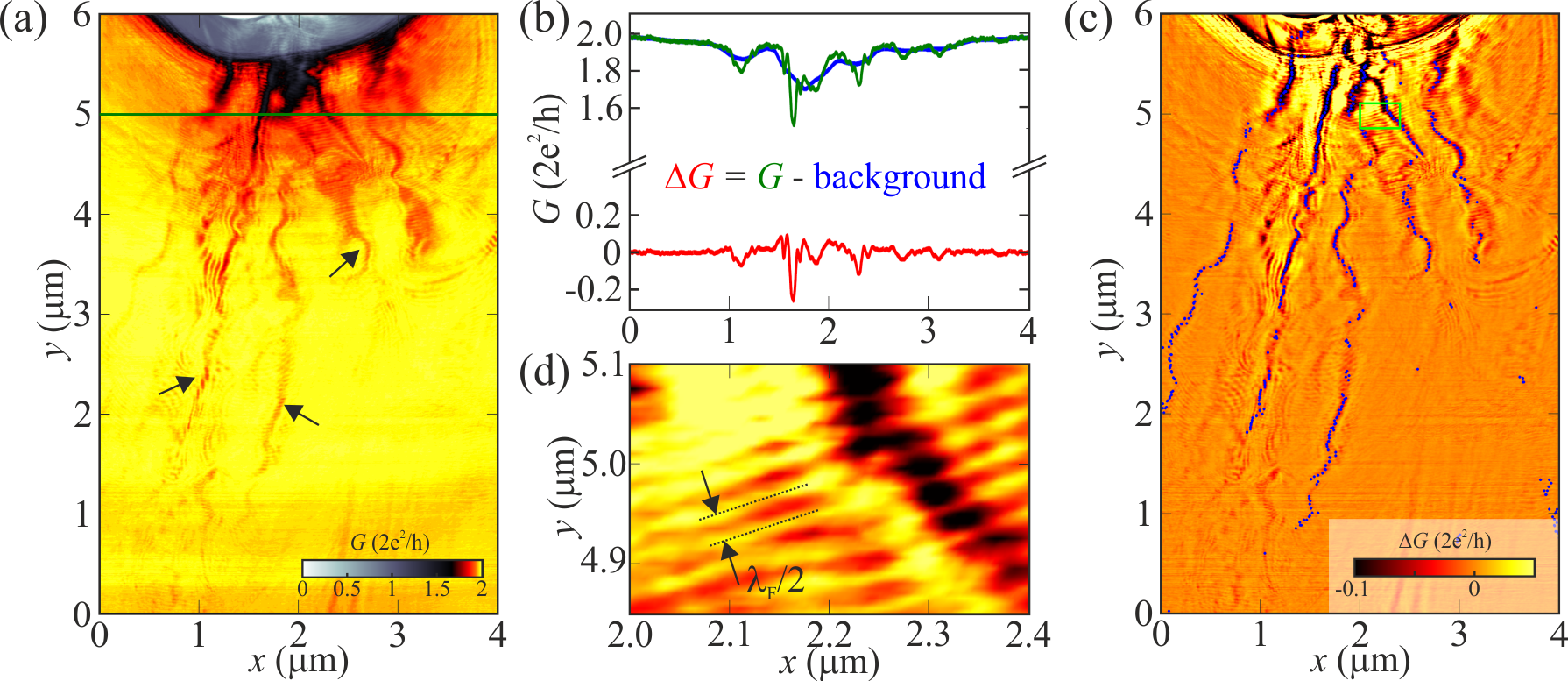}
 \caption{(a) Conductance as a function of tip position displays the pattern of branched electron flow (examples of branches marked by black arrows). Split-gate voltage $V_{\mathrm{QPC}}$ is tuned such that QPC conductance is on second plateau when the tip is in the scan frame center. (b) Conductance variation $\Delta G$ (red) is obtained by subtracting a smoothed background (blue) from the measured conductance (green cut in (a)). (c) $\Delta G$ with minima indicated by blue dots. (d) Conductance variation due to interference effects, position indicated by the green rectangle in (c).}
 \label{fig2label}
 \end{center}
\end{figure}

We tune the QPC to the second conductance plateau at $G = I_{\mathrm{SD}}/V_{\mathrm{L}}=2\times 2e^2/h$ and scan the tip close to the QPC at a distance of 30~nm above the GaAs surface. The tip voltage $V_{\mathrm{tip}}=\SI{-8}{V}$ is chosen such that it creates a disk of zero electron density in the 2DEG. The recorded conductance $G$ as a function of tip position at the highest charge carrier density $n=2.03\times 10^{11} \mathrm{cm}^{-2}$ is presented in Fig.~\ref{fig2label}(a). If the tip is close to the QPC the tails of the tip potential shift the saddle point potential in the constriction and we observe a smooth reduction of $G$. The color scale is chosen such that it depicts the region of strongly reduced conductance in the grey scale and the small variations on the second plateau in the orange scale.
The pattern of branched electron flow is visible (examples are marked by black arrows) but obscured by the smooth background variation of $G$. Therefore we compute the conductance variation $\Delta G$ by subtracting the background from the measured $G$. This data processing is illustrated in Fig.~\ref{fig2label}(b) with the example of $G(x,y=5~\si{\micro\meter})$ marked by the green line in Fig.~\ref{fig2label}(a). The background is calculated by a two-dimensional running average of $G$ with a span of 300~nm in $x$ and 100~nm in $y$. 
The conductance variation $\Delta G$ as a function of tip position is shown in Fig.~\ref{fig2label}(c) with blue points marking local minima, which will be used to compare branch positions at different electron densities. To remove outliers, only local minima with other minima in their vicinity are shown. At $x \approx 2$~\si{\micro\meter}, $y\approx 5.5$~\si{\micro\meter} the transition from the second to the first conductance plateau due to tip gating of the QPC is visible.

Performing the same experiment on the first or third QPC plateau produces similar results. On this sample, the visibility of the branches on the first plateau was lower than on the second. The third plateau is present only in a smaller range of electron density, therefore we used the second plateau for this study.

Thanks to coherent transport through the structure and the high spatial resolution of SGM we also observe interference effects of tip reflected electrons \cite{topinka_coherent_2001}. As an example, Fig.~\ref{fig2label}(d) shows an enlarged view of the area indicated by the green rectangle in Fig.~\ref{fig2label}(c) where the interference fringes are measured. The expected periodicity of half the Fermi-wavelength is indicated by the dotted lines. From earlier SGM studies with tunable electron density \cite{leroy_imaging_2003-1,leroy_imaging_2002,leroy_imaging_2003} it is known that such pattern scale with charge carrier density as expected. In Fig.~\ref{fig2label}(c) further periodic patterns are visible, for example at $x=\SI{0.8}{\micro\meter}$, $y=\SI{2.3}{\micro\meter}$ or at $x=\SI{3.4}{\micro\meter}$, $y=\SI{3.3}{\micro\meter}$. A possible origin of such patterns is the presence of hard scatterers \cite{kolasinski_interference_2016}.

\section{Dependence on charge carrier density}

\begin{figure}
\begin{center}
 \includegraphics[width=0.8\linewidth]{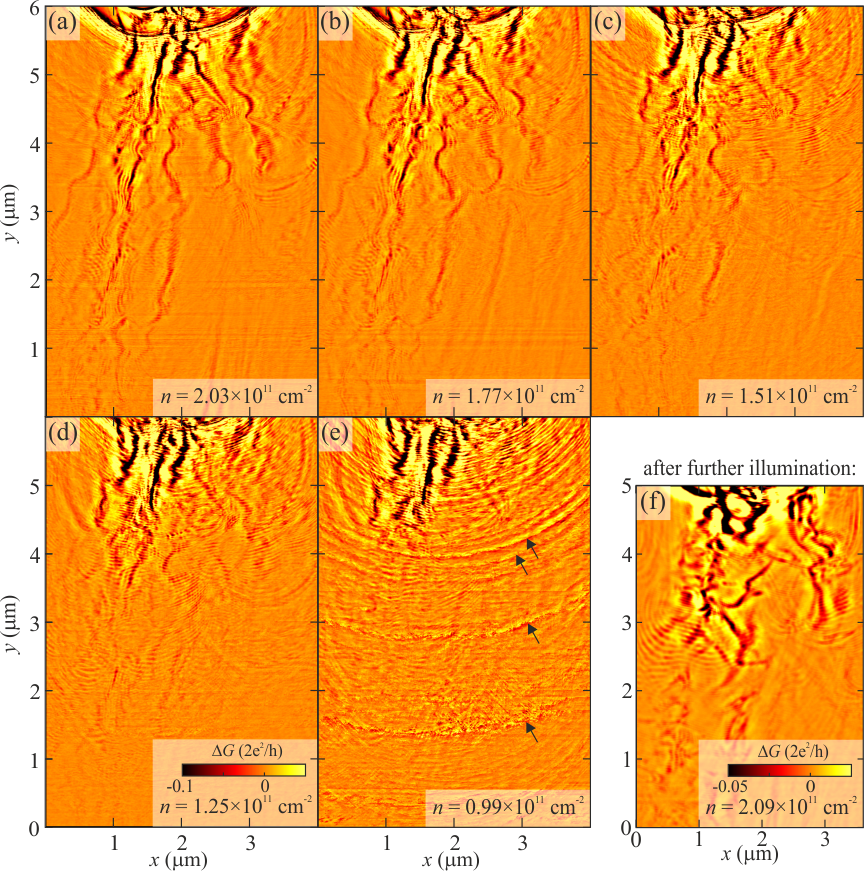}
 \caption{(a)-(e) Conductance variation $\Delta G$ for decreasing electron densities. The pattern of branched flow does not change position as long as it can be detected. Rings of charge rearrangements are marked by black arrows in (e). (f) After additional illumination of the sample, the pattern of the branched flow is different than in (a), while electron density and mobility are unchanged.}
 \label{fig3label}
 \end{center}
\end{figure}

We repeat the measurement of $G$ as function of tip position at four different $V_{\mathrm{bg}}$ leading to lower electron densities. 
To keep the diameter of the depletion disk below the tip constant at all $n$, the amplitude of the tip-induced potential in the 2DEG should grow proportionally to the Fermi energy. We repeat scans with varying $V_\mathrm{tip}$ at different $n$ (data not shown) to find the least negative $V_\mathrm{tip}(V_\mathrm{bg})$ at which branches are visible. This critical tip voltage corresponds to the creation of a depletion disk, \textit{i.e.} when the maximum of the tip induced potential equals the Fermi energy. For all presented measurements we set the tip voltage $V_\mathrm{tip}=-1.42\times V_\mathrm{bg} - \SI{6.87}{V}$ that lies below the critical tip voltage and therefore creates a finite size depletion disk. From the measured $G(x,y)$ at all densities we subtract a smoothed background as described above to calculate the conductance variation $\Delta G$. The results in Figs.~\ref{fig3label}(a)-(c) show that the pattern of branched flow is only weakly modified when reducing the electron density by 25~percent. However, at even lower $n$ and $\mu$ we see in Figs.~\ref{fig3label}(d)-(e) that the region, where branched electron flow is observed, is limited to the vicinity of the QPC. 

In Fig.~\ref{fig3label}(e) we observe rings of charge rearrangements in $\Delta G$, the most prominent rings are marked by black arrows. They occur because the tails of the tip potential rearrange trapped discrete charges. Such a change in its electronic environment shifts the QPC conductance curve with respect to the applied top-gate voltage. Therefore $G$ is not affected by charge rearrangements if the QPC is on a conductance plateau. At low electron densities, the QPC plateaus are tilted as seen in Fig.~\ref{fig1label}(c) and therefore the rings of charge rearrangements are visible in Fig.~\ref{fig3label}(e).

\begin{figure}
\begin{center}
 \includegraphics[width=0.7\linewidth]{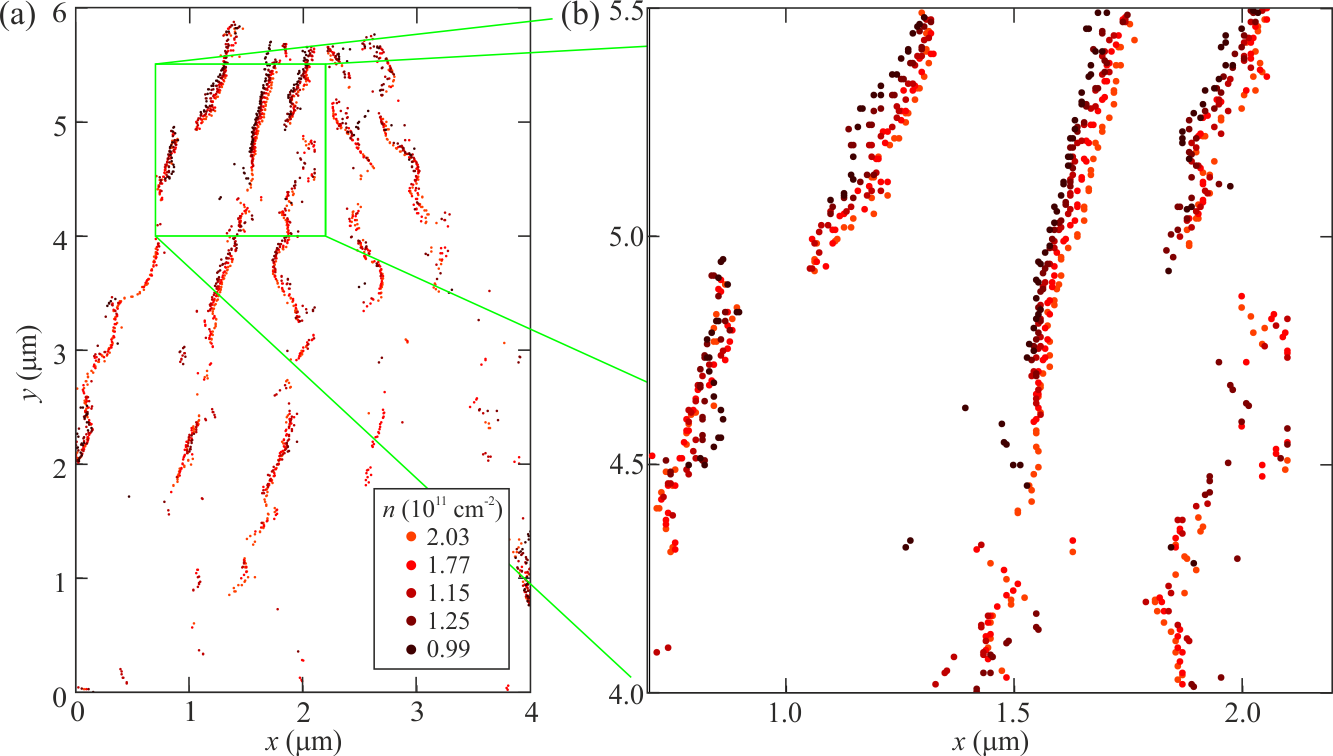}
 \caption{(a) Minima of $\Delta G$ for different charge carrier densities $n$ as described in the main text. (b) Enlarged view shows minor shifts of minimum positions with~$n$.}
 \label{fig4label}
 \end{center}
\end{figure}

To compare the branching pattern at different charge carrier densities, we track the branches by finding local minima in $\Delta G$ as shown in Fig.~\ref{fig2label}(c). The positions of the minima are plotted in Fig.~\ref{fig4label}(a) with different colors indicating their respective charge carrier density. This analysis shows that even when changing the charge carrier density by a factor of two, the branching pattern remains roughly at the same position and the detailed features of the pattern are largely independent of $n$. Figure~\ref{fig4label}(b) is an enlarged view of the square indicated in Fig.~\ref{fig4label}(a) to show that there are systematic minor shifts of the minima of the order of 100~nm whereas the overall pattern remains unchanged.

For comparison, Fig.~\ref{fig3label}(f) shows the pattern of branched electron flow after additional illumination of the sample with respect to the measurements in Figs.~\ref{fig3label}(a)-(e). The additional illumination was short enough to change the charge carrier density by only 3\% at the same back-gate voltage, and no measureable difference in mobility was found (data not shown). Comparing the two measurements in Figs.~\ref{fig3label}(a) and (f) demonstrates that the microscopic pattern of branched electron flow has changed completely due to a significant change in the background potential. Measurements of the macroscopic quantities $n$ and $\mu$ overlook this modification.

\section{Trajectory simulations}
The stability of the distribution of electron motion in a background potential has been studied within previous theoretical work. For example, Liu and Heller \cite{liu_stability_2013} used a fully quantum mechanical model to show that the pattern is robust against changes of the injection into the 2DEG. We will use a less involved, classical model to investigate the stability against changes of the electron density and the Fermi energy. Such trajectory simulations have been used to model branched electron flow in the past \cite{topinka_coherent_2001,heller_branching_2003,steinacher_scanning_2016}. Can they also explain the experimentally observed stability, \textit{i.e.} does a classical particle follow the same trajectory in a random potential if it's kinetic energy changes by a factor of two? To answer this question, we use a trajectory model similar to Steinacher \textit{et al.} \cite{steinacher_scanning_2016}. We model the potential in the 2DEG caused by SGM tip and QPC gates by calculating the charge distribution in Thomas-Fermi approximation with the finite element software \emph{COMSOL 5.0}. The three-dimensional modelling of the sample and the SGM tip includes the screening effects of the top-gates on the tip potential.

\begin{figure}
\begin{center}
 \includegraphics[width=0.9\linewidth]{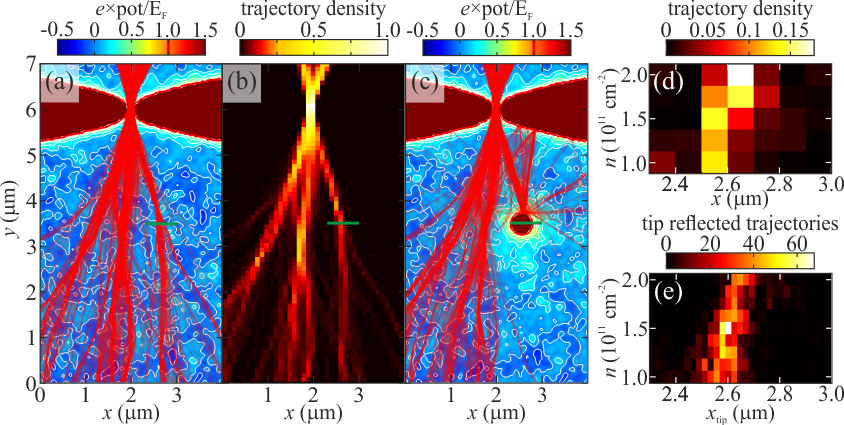}
 \caption{Classical modelling of electron trajectories: (a) Transmitted electron trajectories (red) and potential of QPC and remote donors indicated by the color scale. (b) Normalized number of trajectories in (a) per \SIlist[list-pair-separator={$\times$}]{100;100}{\nano\metre} square. (c) Transmitted trajectories in a potential including the voltage biased tip at $x_{\mathrm{tip}}=2.56$~\si{\micro\meter}, $y_{\mathrm{tip}}=3.5$~\si{\micro\meter}. (d) Trajectory density without tip perturbation at the position of the green line in (b) for different charge carrier densities. (e) Number of electron trajectories, that are reflected at the tip and backscattered through the QPC, for tip positions and charge carrier densities as in (d).}
 \label{fig5label}
 \end{center}
\end{figure}

We add a correlated random background potential to the solution of the \emph{COMSOL} simulation. It includes Thomas-Fermi screening of the donor atoms \cite{ando_electronic_1982}, the thickness-dependence of the 2DEG on the back-gate voltage \cite{ando_electronic_1982,fang_negative_1966,davies_physics_1997}, and correlation of ionized donors \cite{efros_maximum_1990}.
The correlation parameter of the donors is used to adjust the calculated charge carrier mobility to the experimental values. For comparison, both are shown in Fig.~\ref{fig1label}(b). The sum of random background potential and \emph{COMSOL} simulation is shown as color map in Fig.~\ref{fig5label}(a) without SGM tip and in Fig.~\ref{fig5label}(c) in the presence of the tip. To simulate the motion of electrons in the potential, we calculate trajectories starting at $y=7.5$~\si{\micro\meter} from an equidistant grid of $x$-positions and angles with a kinetic energy equal to the Fermi energy. Only the transmitted trajectories are added to Figs.~\ref{fig5label}(a) and (c) as red lines. They show the uneven spatial distribution and caustics that are expected for trajectories in a correlated potential \cite{heller_branching_2003}. 

How are these simulations without tip potential related to the SGM experiment? Figure~\ref{fig5label}(b) shows the normalized number of trajectories per \SIlist[list-pair-separator={$\times$}]{100;100}{\nano\metre} square of the simulation in (a). The features qualitatively agree with the measured $\Delta G$. We repeat the simulation for the values of $n$ measured in the experiment. Figure~\ref{fig5label}(d) illustrates the evolution of the trajectory density as a function of $n$ along the green line in Fig.~\ref{fig5label}(b) ($x=2.3-3.0$~\si{\micro\meter}, $y=3.5$~\si{\micro\meter}). 

To directly compare the model to the experiment, we also simulate the situation with a SGM tip potential at different positions $x_{\mathrm{tip}}$, $y_{\mathrm{tip}}$ for different $n$. Trajectories, which are transmitted through the QPC, reflected by the tip potential, and returning through the QPC, will be called \emph{tip reflected trajectories}. They are experimentally relevant because they contribute to the sample resistance by being backscattered through the constriction. The number of tip reflected trajectories is shown in Fig.~\ref{fig5label}(e) for the same tip coordinates $x_{\mathrm{tip}}$, $y_{\mathrm{tip}}$ as the positions $x$, $y$ without tip in Fig.~\ref{fig5label}(d). The ratio of tip reflected to transmitted trajectories is up to 10\%, similar to the experimentally observed ratio $\Delta G/G$. The similarity of Figs.~\ref{fig5label}(d) and (e) confirms the findings of Topinka \textit{et al.}: The ratio of tip reflected trajectories to the total number of trajectories hitting the tip head-on is roughly constant. Therefore the number of tip reflected trajectories maps the local trajectory density in the absence of the tip \cite{topinka_imaging_2000,heller_branching_2003}, even though the reflection of trajectories happens at the edge of the tip depleted region and not at its center. The experimentally observed minor shifts of branch position as a function of electron density shown in Fig.~\ref{fig4label}(b) are reproduced in both simulations with and without tip potential. In Figs.~\ref{fig5label}(d) and (e) we see this effect as the shift of the maxima with $n$.

\begin{figure}
\begin{center}
 \includegraphics[width=0.9\linewidth]{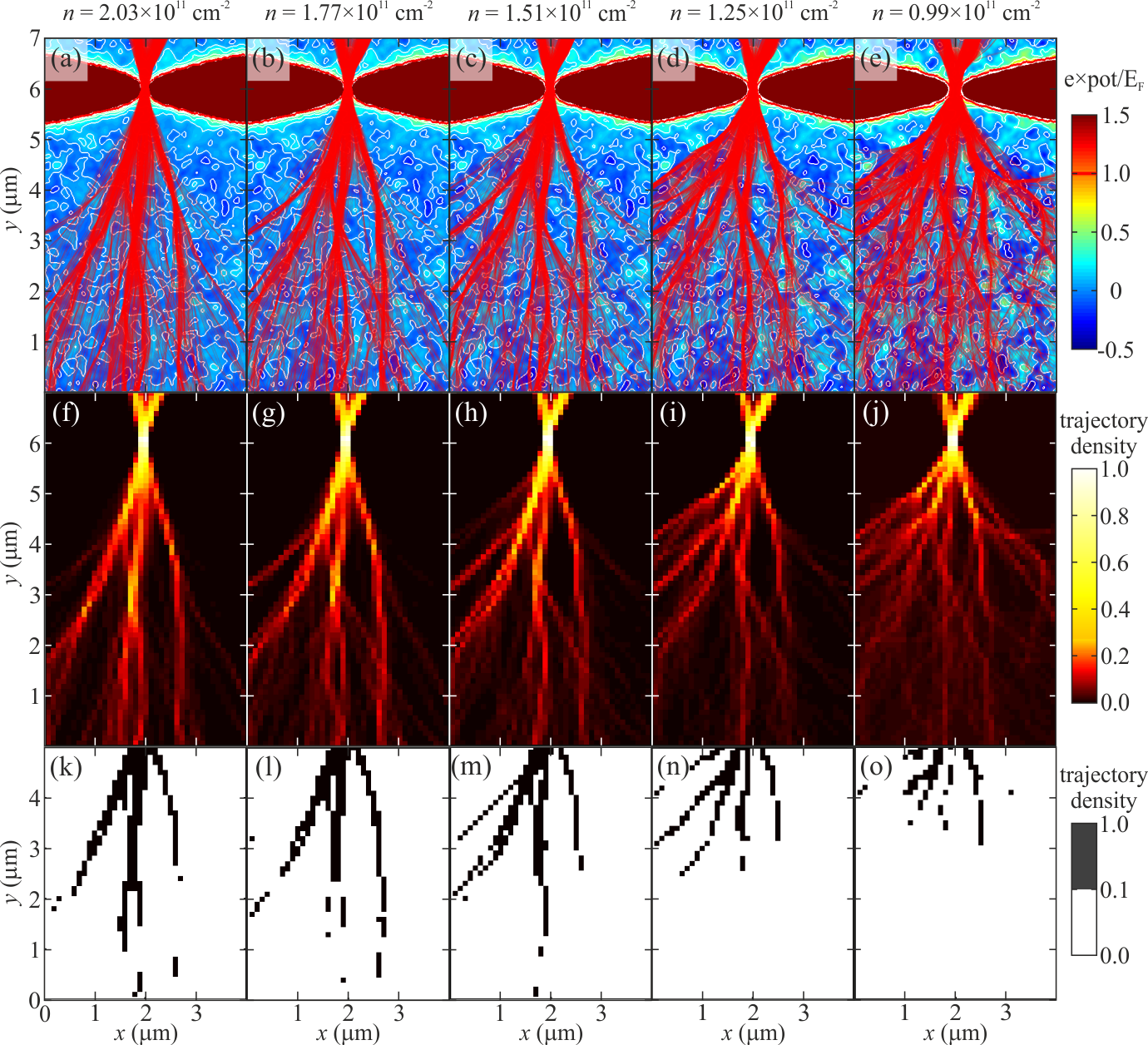}
 \caption{Classical modelling at different electron densities: (a)-(e) Transmitted electron trajectories (red) and potential of QPC and remote donors indicated by the color scale. (f)-(j) Normalized number of trajectories in (a)-(e) per \SIlist[list-pair-separator={$\times$}]{100;100}{\nano\metre} square. With decreasing $n$ the regions of high density are limited to the vicinity of the QPC. (k)-(o) Experimentally accessible region of (f)-(j) drawn black if trajectory density is above the threshold value~0.1. Those pattern are robust against changes in~$n$.}
 \label{fig6label}
 \end{center}
\end{figure}

In Fig.~\ref{fig6label} simulations for the same densities as in the measured $\Delta G$ of Figs.~\ref{fig3label}(a)-(e) are presented. The trajectories in Figs.~\ref{fig6label}(a)-(e) show increasingly complex behavior with decreasing $n$. This can be attributed to the reduced mobility, as the background potential is higher with respect to the Fermi energy. 
We calculate the trajectory densities in Figs.~\ref{fig6label}(f)-(j) as described above, which are related to the number of tip reflected trajectories and therefore to the reduced conductance in the experiment. A certain signal strength is required to determine the pattern experimentally, so a threshold of the relative trajectory density is chosen at an arbitrary value of 0.1. Figures~\ref{fig6label}(k)-(o) show the trajectory density above this threshold value in black for the experimentally accessible region in front of the QPC. 

The maps of trajectory density above threshold in Figs.~\ref{fig6label}(k)-(o) reproduce the features of the measurements in Figs.~\ref{fig3label}(a)-(e) qualitatively: The branched flow pattern remains overall constant as long as it is visible. 
To observe the pattern far from the QPC, a high $n$ and $\mu$ is required, otherwise the electron trajectories are no longer focussed in the caustics but are evenly spread in the 2DEG. In the experiment, the pattern is lost because the number of electrons scattering from the tip potential is the same at every tip position. This effect can be described by choosing a threshold of 0.1 in trajectory density.

Does the observed stability of trajectory density also hold for the behavior of the individual trajectories? We found an unstable behavior of the \emph{tip reflected trajectories}: A small difference (below 1~\textperthousand) of the charge carrier density may change a particular tip reflected trajectory to become a transmitted trajectory (from the same starting position and starting direction). A similar behavior is observed by small changes in tip positions which are below the experimental resolution. Both the convex saddle point potential and the background potential make two initially close trajectories to separate after a distance of a few micrometers. 
From this behavior we conclude that single electron trajectories fluctuate within their bundles and the observed stability is due to averaging over multiple trajectories.

\section{Conclusions}
We have experimentally shown that the pattern of branched electron flow is astonishingly robust against changes of the charge carrier density. On the other hand, the branched flow pattern can be changed completely by additional illumination which modifies the background potential without changing macroscopic quantities as charge carrier density or mobility. The effect of such modifications are only visible by local investigations.

Additionally, we have presented a model to simulate the classical motion of electrons in a random potential background. The simulation reproduces the experimentally observed branch stability including minor shifts in branch position. 

\section*{Acknowledgements}
We thank D. Weinmann, R. Jalabert, and O. Ly for fruitfull discussions. The authors acknowledge financial support from ETH Z\"urich and from the Swiss National Science Foundation (SNF 2-77255-14 and NCCR QSIT).

\section*{References}
\bibliography{stable_branch_bib_v1}

\end{document}